%% file: THirota_1.tex
\documentclass{evn2004}
\usepackage{txfonts}
\begin{document}
\include{page}
   \title{Observations of H$_{2}$O maser sources in 
           Orion-Monoceros Molecular Clouds with VERA}

   \author{T. Hirota\inst{1}
          \and
          VERA project team\inst{1,2}
          }

   \institute{National Astronomical Observatory of Japan, 
Osawa 2-21-1, Mitaka-shi, Tokyo 181-8588, Japan
         \and
            Department of Physics, Kagoshima University, 
Korimoto 1-21-35, Kagoshima-shi, Kagoshima 890-0065, Japan
             }

   \abstract{
We present results of phase-referencing VLBI observations of 
H$_{2}$O maser sources in Orion-Monoceros Molecular Clouds with VERA 
(VLBI Exploration of Radio Astrometry), which is newly constructed Japanese 
VLBI network. 
Main topics of this poster are (1) the aim of one of the first scientific projects 
for VERA "3-Dimensional Structure and Kinematics of Orion-Monoceros Molecular 
Cloud Complex"; (2) current status (sensitivity and astrometric accuracy) of 
phase-referencing VLBI observations with VERA; and (3) results of 
VLBI observations of H$_{2}$O maser sources in Orion-Monoceros Molecular 
Clouds with VERA. 
   }

   \titlerunning{Observations of H$_{2}$O maser sources in 
           Orion-Monoceros Molecular Clouds with VERA}
   \authorrunning{T. Hirota et al. }

   \maketitle
%
%________________________________________________________________

\section{Introduction}

VERA is a newly 
constructed Japanese VLBI network dedicated to phase-referencing 
VLBI observations (Kobayashi et al. 2003). 
The main goal of the VERA project is to reveal 3-dimentional 
Galactic structure and kinematics based on the precise astrometric 
VLBI observations of annual parallax and proper motion of 
H$_{2}$O and SiO maser sources with an accuracy of 10$\mu$as. 
For the purpose of this, we at first plan to carry out 
3 projects as test observations of VERA to measure distance to 
(1) the Galactic center, 
(2) Mira-type variables to establish period-luminosity relation, 
and (3) Orion-Monoceros molecular cloud complex, which is the 
nearest giant molecular cloud (GMC) complex. In this poster, 
we introduce details of one of the first 
VERA project observations "3-Dimensional Structure and 
Kinematics of Orion-Monoceros Molecular Cloud Complex". 

Orion-Monoceros molecular cloud complex is the nearest 
GMC at the distance of about 400-800 pc (e.g. Genzel et al. 1981), 
Therefore, it has been recognized as an important object to study 
star-formation processes. 
Using VERA, we aim to measure the distance and proper motion of 
each GMC to reveal kinematics of Orion-Monoceros molecular cloud 
complex. Because more than 20 H$_{2}$O maser sources associated 
with young stellar objects have been detected, 
Orion-Monoceros molecular cloud complex is the best 
target for our systematic VLBI observations.  
In addition, Orion-Monoceros molecular cloud complex is 
located in one of the Galactic arms, Local arm, 
and hence, this project would be the first step 
to complete the 3-D map of our Galaxy. 

\section{Observations}

VLBI observations of H$_{2}$O maser sources 
in Orion-Monoceros molecular cloud complex were  
carried out in several observing sessions in 2003 and 2004. 
All the 4 VERA stations were used in most of observing sessions, 
although only 3 stations were used in part of 
observing sessions due to bad weather condition. 

For phase-referencing observations, 
all the observations were made in the dual beam mode; 
a H$_{2}$O maser source and its reference source with a separation 
angle of 0.3-2.2 degree were observed simultaneously with 
the dual beam receivers on VERA using a dual-beam phase calibration 
based on the horn-on-dish method 
(Kobayashi et al. 2003; Honma et al. 2003). 
Left-handed circular 
polarization was received and sampled with 2-bit quantization, 
and filtered using a digital filter before being recorded onto 
magnetic tapes. The data were recorded at a rate of 128Mbps. 
%(16 MHz bandwidth for each source). 
In several observing sessions, we used wide band recording system 
at a rate of 1Gbps. 
%(16 MHz bandwidth for a H$_{2}$O maser and 240 MHz for a reference source). 
Observations of strong continuum source, J0530+1331, were made every 
1-2 hours for calibration purposes. 
Typical system temperature measured with the chopper-wheel method 
were 100-600 K, depending on weather conditions at each station. 

Correlation process was carried out on Mitaka FX correlator 
located at the NAO Mitaka campus with a spectral resolution of 
15.625kHz (1024 channels per 16MHz band). 
Data reduction was performed in the standard way using the NRAO AIPS. 
In the analysis, 
fringe fitting and self calibration were performed on 
the reference source, and the solutions were applied to 
target maser source. 

\section{Results}

\subsection{A Survey of H$_{2}$O maser sources with VERA}

It is difficult to survey and continue 
long-term monitoring observations of H$_{2}$O maser sources with VLBI, 
especially for low-mass star-forming regions 
because of their variability (Claussen et al. 1996). 
In addition, H$_{2}$O maser sources in nearby molecular clouds 
are heavily resolved out with the VLBI observations 
(e.g. Migenes et al. 1999). 
Therefore, we at first carried out a survey observations of H$_{2}$O 
maser sources in Orion-Monoceros molecular cloud complex 
and their reference sources with VERA. 
Observations were carried out with the VERA 3 stations 
(Mizusawa, Iriki, and Ishigaki) on 2003 October 15 UT16-20h. 
A maser source and a continuum source were observed simultaneously, 
and the data were recorded at a rate of 128 Mbps 
(16 MHz bandwidth for each source). 

We observed 15 H$_{2}$O maser sources and 11 continuum 
sources. Integration time for each source is 10 minutes. 
Table \ref{tab-survey} shows the observed maser sources. 
Only 5 sources were detected with one of the baselines of VERA. 
Low detection rate of H$_{2}$O maser sources is mainly due to 
their variability (Claussen et al. 1996), while 
it is also likely that 
some of the maser sources (e.g. Mon R2) were completely resolved out 
even with the shortest baseline of VERA (1000 km). 
For continuum sources, we detected only 2 known sources, J0541-0541 and 
J0607-0834. In order to detect more reference sources, we are 
planning to carry out further VLBI observations at a rate of 1Gbps. 

\begin{table}[th]
\begin{footnotesize}
\begin{center}
\caption{Observed H$_{2}$O maser sources}
\label{tab-survey}
\begin{tabular}{lrrl}
& & \\
\hline
H$_{2}$O maser      & \hspace{10pt} $F_{tot}^{\mathrm{a}}$   &
   \hspace{10pt} $F_{cor}^{\mathrm{b}}$ & Reference \\
Source  & (Jy)        & (Jy)    & Source    \\
\hline
Orion KL        &  2500 & 600  & J0541-0541 \\
OriA-W          &     5 & $<$2 & NO         \\
OMC-2           &    40 &   6  & J0541-0541 \\
HH1             &    60 &  16  & J0541-0541 \\
L1641-MMS1      &  $<$2 & $<$2 & J0541-0541 \\
NGC2024 FIR5    &    15 & $<$2 & NO         \\
IRAS 05413-0104 &  $<$2 & $<$2 & NO         \\
NGC2071         &   200 &  50  & NO         \\
NGC2071N        &  $<$2 & $<$2 & NO         \\
IRAS 05393-0156 &    15 & $<$2 & NO         \\
IRAS 05445+0016 &     6 & $<$2 & NO         \\
B35             &  $<$2 & $<$2 & NO         \\
Mon R2 IRS3     &   150 & $<$2 & J0607-0834 \\
HH12-15         &    90 &  60  & NO         \\
HH19-27         &  $<$2 & $<$2 & NO         \\
\hline
\multicolumn{4}{l}{$^{\mathrm{a}}$ Total power flux observed with the 20 m antenna. } \\
\multicolumn{4}{l}{$^{\mathrm{b}}$ Correlated flux observed with the VERA Mizusawa-} \\
\multicolumn{4}{l}{ Iriki baseline. } \\
\end{tabular}
\end{center}
\end{footnotesize}
\end{table}

\subsection{Monitoring of H$_{2}$O Masers Sources with VERA}

According to our survey observations, 5 sources 
(Orion KL, OMC-2, HH1, NGC2071, and HH12-15) were detected with 
VERA. Among them, 3 sources (Orion KL, OMC-2, and HH1) 
are associated with the reference source (J0541-0541) 
within 2.2 degree and hence, they would be good candidates for 
monitoring observations with VERA. 
In addition, intense H$_{2}$O maser was 
detected toward Mon R2 in the total power spectra. Since Mon R2 
was detected with previous VLBI observations (Migenes et al. 1999), 
it is also a possible candidate for our 
project. 

Therefore, we have started to carry out monitoring observations 
of 4 maser sources (Orion KL, OMC-2, HH1, and Mon R2) 
with VERA since 2004 January. 
Average interval of observation is 1 month. 
The total observing time was 4 hours for each 
source including calibration. 
At present, 7 epochs of observations have been made and 
correlation processes have been done for 4 epochs. 

\begin{figure}
\centering
%\vspace{307pt}
\vspace{270pt}
\includegraphics{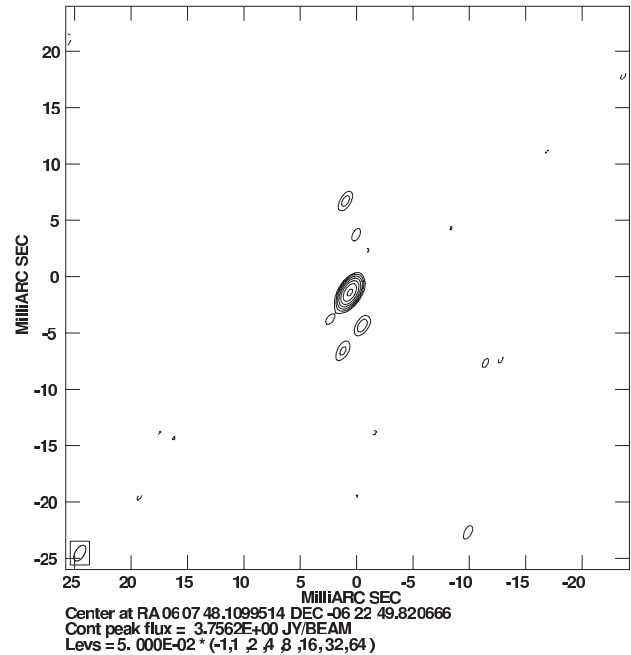}
\caption{A phase-referenced image of H$_{2}$O maser sources 
in Mon R2. The LSR velocity is 5.9 km s$^{-1}$.  \label{fig-map}}
\end{figure}

Figure \ref{fig-map} shows an example of 
a phase-referenced image of the H$_{2}$O maser source Mon R2 
observed with VERA. A natural weighted map was produced using 
the AIPS task "IMAGR". The synthesized beam was 1.6$\times$0.8 
mas with the position angle of -31 degree. 
In this epoch, another maser feature (11.4 km s$^{-1}$) is also 
detected in Mon R2, although it seems to be resolved out. 
On the other hand, we could not detect the same maser feature at 
the LSR velocity of 5.9 km s$^{-1}$ in 
other 3 epochs. 
In order to detect annual parallax and proper motion, 
further observations and data analysis including 
other sources (Orion KL, OMC-2, and HH1) are now in progress. 

\begin{acknowledgements}
TH is financially supported by Grant-in-Aid from Ministry of 
Education, Science, Sports and Culture of Japan 
(13640242 and 16540224).
\end{acknowledgements}

\end{document}

%% file: page.tex
\setcounter{page}{201}

%% file: THirota_1.bbl
\begin{thebibliography}{}
%
\bibitem[1996]{claussen} Claussen, M. J., Wilking, B. A., Benson, P. J., 
  Wootten, A., Myers, P. C., \& Terebey, S. 1996, ApJS, 106, 111
\bibitem[1981]{genzel} Genzel, R., Reid, M. J., Moran, J. M., \& 
      Downes, D. 1981, ApJ, 244, 884 
\bibitem[2003]{honma1} Honma, M. et al. 2003, PASJ, 55, L57
%\bibitem[2004]{honma2} Honma, M. et al. 2004, PASJ, 56, L15
\bibitem[2003]{kobayashi} Kobayashi, H. et al. 2003, in ASP Conf. Ser. 306, 
      New technologies in VLBI, ed. Y.C. Minh, 367
\bibitem[1999]{migenes} Migenes, V. et al. 1999, ApJS, 123, 487
%
\end{thebibliography}
